\documentclass[12pt,nofootinbib,superscriptaddress]{revtex4}

\usepackage{amssymb,color,paralist,amsmath,epsfig}
\usepackage{graphicx}
\usepackage{amsfonts}
\usepackage{nicefrac,esint}
\usepackage{verbatim}
\usepackage{float}
\usepackage{verbatim}

\usepackage{relsize}
 
\usepackage[colorlinks,citecolor=blue,linktoc=all,linkcolor=cyan]{hyperref}
\usepackage[section]{placeins}

\newcommand{\sla}[1]{{\not\! #1}} 

\newcommand{\beq}{\begin{equation}}
\newcommand{\eeq}{\end{equation}}

\newcommand{\ber}{\begin{eqnarray}} 
\newcommand{\eer}{\end{eqnarray}}

\usepackage{color}

\usepackage[normalem]{ulem}

\renewcommand\sout{\bgroup \color[rgb]{0,0.00,1.} \ULdepth=-.5ex \ULset}

\begin{document}

\title{Soft-photon corrections to the Bethe-Heitler process in the $\gamma p\rightarrow l^+l^-p$ reaction}
\author{Matthias Heller}
\affiliation{Institut f\"ur Kernphysik and PRISMA Cluster of Excellence, Johannes Gutenberg Universit\"at, Mainz, Germany}
\author{Oleksandr Tomalak}
\affiliation{Institut f\"ur Kernphysik and PRISMA Cluster of Excellence, Johannes Gutenberg Universit\"at, Mainz, Germany}
\author{Marc Vanderhaeghen}
\affiliation{Institut f\"ur Kernphysik and PRISMA Cluster of Excellence, Johannes Gutenberg Universit\"at, Mainz, Germany}

\date{\today}

\begin{abstract}
We report on the calculation of first-order QED corrections for the $\gamma p\rightarrow l^+l^-p^{\prime}$ process. An upcoming experiment at MAMI (Mainz) aims to compare the cross sections of muon- and electron-pair production in this reaction to test lepton universality. Precise knowledge of the electromagnetic radiative corrections is needed for these measurements. As a first step, we present the leading QED radiative corrections in the soft-photon approximation when accounting for the finite lepton mass. For the kinematics at MAMI, we find corrections of the percent level for muons, and of order $10\%$ for electrons. 

\end{abstract}

\maketitle

\tableofcontents

\vspace{-0.2cm}

\section{Introduction}
Recent experiments found a significant difference in the proton charge radius, comparing measurements with electrons and muons. Currently, the most precise measurements with electron scattering were performed by the A1 Collaboration in Mainz \cite{Bernauer:2010wm,Bernauer:2013tpr}. The proton radius extracted from these measurements is  $ R_E = 0.879(8) ~\mathrm{fm}$. For muonic measurements, there is currently only the proton radius extraction by muonic spectroscopy \cite{Pohl:2010zza,Antognini:1900ns}, yielding a significantly smaller value than the extraction by electron scattering experiments. The reported value of the muonic hydrogen experiments is $ R_E = 0.84087(39) ~\mathrm{fm}$ \cite{Antognini:1900ns}.

This discrepancy, often referred to as the proton radius puzzle, has triggered a lot of activity in recent years. Explanations for this puzzle reach from systematic errors in the extraction, see Refs. \cite{Lorenz:2012tm,Lorenz:2014vha,Lorenz:2014yda,Lee:2015jqa,Arrington:2015ria,Arrington:2015yxa,Griffioen:2015hta,Higinbotham:2015rja}, to physics beyond the standard model, see for example in Refs. \cite{TuckerSmith:2010ra,Barger:2010aj,Barger:2011mt,Batell:2011qq,Brax:2010gp,Jentschura:2010ha,Carlson:2012pc,Wang:2013fma,Onofrio:2013fea,Karshenboim:2014tka}. If one tries to explain this puzzle by new physics, one has to give up lepton universality as a consequence, since this requires the same, universal coupling for all leptons. 

To shed further light on this puzzle and test lepton universality, the MUSE experiment has been proposed, which aims at comparing the scattering of muons and electrons on a proton target \cite{Gilman:2013eiv,Gilman:2017hdr}. In Ref. \cite{Pauk:2015oaa}, the authors suggested another test of lepton universality by comparing the cross section of lepton-pair production for muons and electrons in the process $\gamma\; p\rightarrow l^+l^-p^{\prime}$. Such experiment only requires a relative measurement through the ratio of electron- and muon-pair production cross sections slightly above di-muon production threshold. According to the finding of Ref. \cite{Pauk:2015oaa}, the measurement of this ratio with absolute precision of around $7\times 10^{-4}$ can test lepton universality at $3\sigma$ significance level. An upcoming experiment at MAMI is planned to perform such measurements \cite{MAMI_photopr}.

For a precise theoretical prediction, it is, however, necessary to include higher-order corrections to the process. In this article, we report as a first step on the calculation of the first-order QED corrections in the soft-photon limit when accounting for the finite lepton mass.

The outline of the paper is as follows. In Sec. \ref{sec2}, we introduce the kinematical notations for the process $\gamma p\rightarrow l^+l^- p^{\prime}$ and give the formulas for the cross section at tree level. In Sec. \ref{sec3}, we evaluate the first-order QED corrections to the cross section in the soft-photon approximation. This limit is defined by a soft scaling of the loop momenta. We give the analytic expressions for the real and virtual corrections. We show that they factorize in terms of the tree level cross section, and explicitly check the cancellation of infrared divergences. In Sec. \ref{sec4}, we present the results of this work. We quantify how the ratio of cross sections of muon- and electron-pair production to electron-pair production is affected by radiative corrections. We give our conclusions and an outlook in Sec. \ref{sec5}.

\label{sec1}
\section{Lepton-pair production at tree level}
\label{sec2}
\begin{figure}
	\includegraphics[scale=0.70]{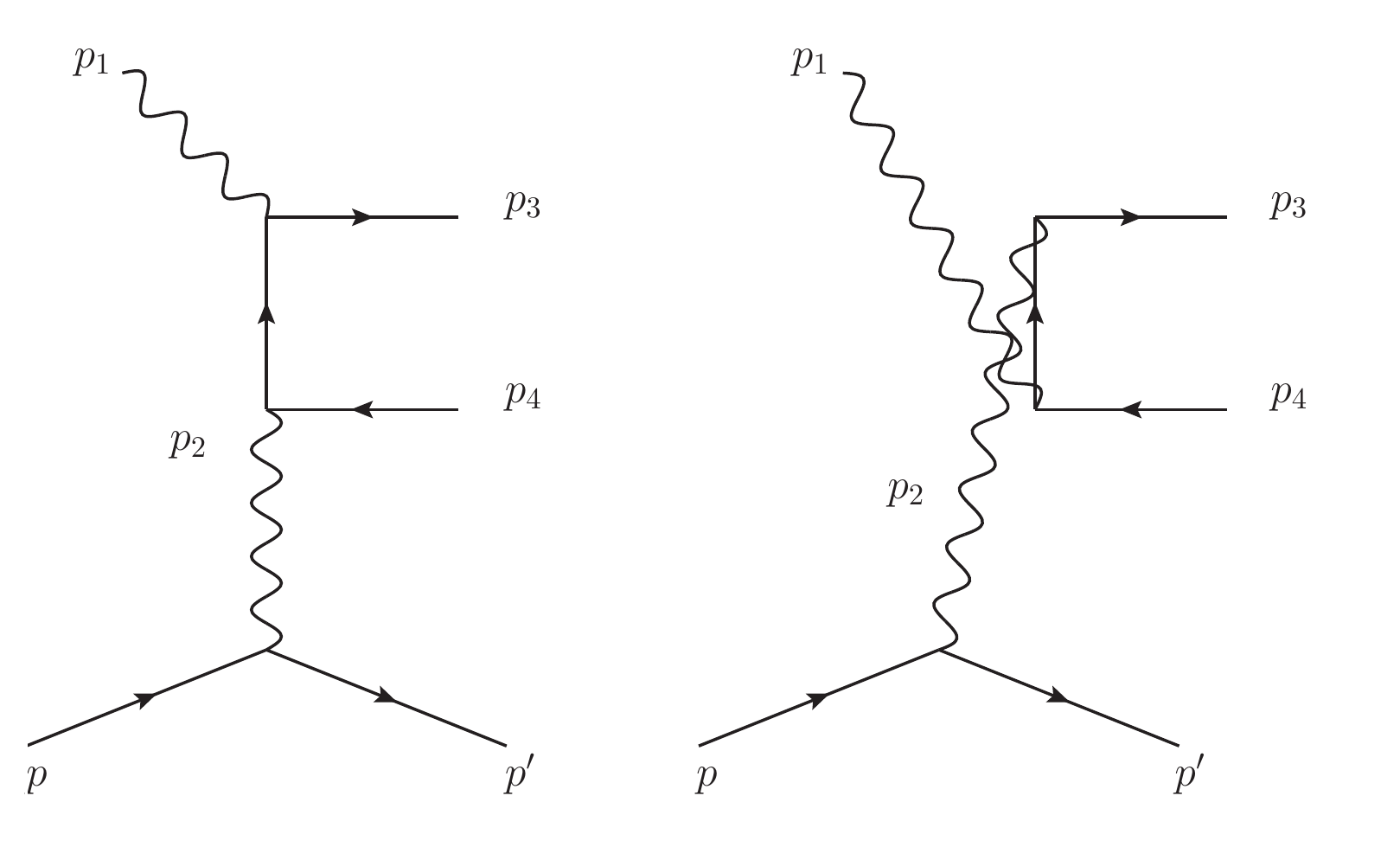}\hspace{0.cm}
	\caption{The Bethe-Heitler process at tree level.\label{tree}}
\end{figure}
The Bethe-Heitler process at tree level is described by two graphs, see Fig. \ref{tree}. We use $p$ ($p^{\prime}$) for the momenta of the initial (final) proton, and $p_3$ ($p_4$) for the momenta of leptons $l^-$ ($l^+$) respectively. The initial photon has momentum $p_1$, and the virtual photon momentum in the one-photon exchange graphs of Fig. \ref{tree} is defined as $p_2=p-p^{\prime}$. The Mandelstam variables for this process are defined as
\begin{align}
(p_3+p_4)^2&=s_{ll},\\
(p_3-p_1)^2&=t_{ll},\\
p_2^2=(p-p^{\prime})^2&=t.
\end{align}
The on-shell condition for external particles implies:
\begin{align}
p_3^2&=p_4^2=m^2,\\
p^2&=p^{\prime\; 2}=M^2,\\
p_1^2&=0.
\end{align}

At leading order, the scattering amplitude $\mathcal{M}_0$ is given by
\begin{align}
\mathcal{M}_0&=\bar{u}(p_3)(ie)\left[\gamma^{\nu}\frac{i(\sla{p}_3-\sla{p}_1+m)}{(p_3-p_1)^2-m^2}\gamma^{\mu}+\gamma^{\mu}\frac{i(\sla{p}_1-\sla{p}_4+m)}{(p_1-p_4)^2-m^2}\gamma^{\nu}\right](ie)v(p_4)\times\nonumber\\
&\times\frac{-i}{t}\varepsilon_{\nu}(p_1)\bar{u}(p^{\prime})(-ie)\Gamma_{\mu}(t)u(p),
\end{align}
where the electromagnetic vertex $\Gamma^{\mu}$ for the proton is expressed as
\begin{equation}
\Gamma^{\mu}(t)=F_D(t)\gamma^{\mu}-iF_P(t)\frac{\sigma^{\mu\nu}(p_2)_{\nu}}{2M},\label{protonVertex}
\end{equation} 
with the proton's Dirac and Pauli form factors $F_D$ and $F_P$, respectively.

The corresponding unpolarized differential cross section $d\sigma_0$ is given by
\begin{equation}
\left(\frac{d\sigma}{dt\,ds_{ll}\,d\Omega_{ll}^{CM_{l^+l^-}}}\right)_0=\frac{1}{(2\pi)^4}\frac{1}{64}\frac{\beta}{(2M\, E_{\gamma})^2}
\left[\overline{\sum_{i}}\sum_{f}\left(\mathcal{M}_0^{*}\;\mathcal{M}_0\right)\right],\label{CrossSectionSum}
\end{equation}
where $E_{\gamma}$ is the lab energy of the initial photon and $\Omega_{ll}^{CM_{l^+l^-}}$ is the solid angle of the lepton pair in their center-of-mass frame, in which the lepton velocity is denoted by
\begin{equation}
\beta=\sqrt{1-\frac{4m^2}{s_{ll}}}.
\end{equation}
In Eq. \eqref{CrossSectionSum}, we average over all polarizations in the initial state and sum over the polarizations in the final state. We express the cross section as a product of hadronic and leptonic parts as
\begin{equation}\label{CrossSection}
\left(\frac{d\sigma}{dt\,ds_{ll}\,d\Omega_{ll}^{CM_{l^+l^-}}}\right)_0=\frac{\alpha^3\beta}{16\pi(2M E_{\gamma})^2\;t^2}L_{\mu\nu}H^{\mu\nu},
\end{equation}
where the fine-structure constant is defined as $\alpha\equiv e^2/4\pi\approx 1/137$. Furthermore, the unpolarized leptonic tensor $L_{\mu\nu}$ (including the average over the initial photon polarization) is given by
\begin{align}
L^{\mu\nu}=-\frac{1}{2}\;\text{Tr}&\left[(\sla{p_3}+m)\left(\gamma^{\alpha}\frac{(\sla{p_3}-\sla{p_1}+m)}{(p_3-p_1)^2-m^2}\gamma^{\mu}+\gamma^{\mu}\frac{(\sla{p_1}-\sla{p_4}+m)}{(p_1-p_4)^2-m^2}\gamma^{\alpha}\right)\right.\nonumber\\
&\left.(\sla{p_4}-m)\left(\gamma^{\nu}\frac{(\sla{p_3}-\sla{p_1}+m)}{(p_3-p_1)^2-m^2}\gamma_{\alpha}+\gamma_{\alpha}\frac{(\sla{p_1}-\sla{p_4}+m)}{(p_1-p_4)^2-m^2}\gamma^{\nu}\right)\right],
\end{align}
and the unpolarized hadronic tensor $H^{\mu\nu}$ by
\begin{equation}
H^{\mu\nu}=\frac{1}{2}\text{Tr}\left[(\sla{p^{\prime}}+M)\;\Gamma^{\mu}\;(\sla{p}+M)\;(\Gamma^{\dagger})^{\nu}\right].
\end{equation}
Using \eqref{protonVertex}, the unpolarized hadronic tensor can be expressed as
\begin{align}
H^{\mu\nu}=(-g^{\mu\nu}+\frac{p_2^{\mu}p_2^{\nu}}{p_2^2})\left[4M^2\tau G_M^2(t)\right]+\tilde{p}^{\mu}\tilde{p}^{\nu}\frac{4}{1+\tau}\left[G_E^2(t)+\tau G_M^2(t)\right],
\end{align}
where $\tilde{p}\equiv (p+p^{\prime})/2$, $\tau\equiv -t/(4M^2)$, and where we conveniently express the hadronic tensor in terms of electric $(G_E)$ and magnetic $(G_M)$ form factors defined as
\begin{align}
G_E&=F_D-\tau F_P,\\
G_M&=F_D+F_P,
\end{align}
which are functions of the spacelike momentum transfer $t$.

For the electric and magnetic proton form factors, which enter the total cross sections for lepton-pair production, we exploit the fit of Ref. \cite{Bernauer:2013tpr}, which is based on a global analysis of the electron-proton scattering data at $Q^2 < 10~\mathrm{GeV}^2$ with an empirical account of TPE corrections.

In the experimental setup, when only the recoil proton is measured, one has to integrate \eqref{CrossSection} over the lepton angles:
\begin{equation}
\left(\frac{d\sigma}{dt\,ds_{ll}}\right)_0=\frac{\alpha^3\beta}{16\pi (2ME_{\gamma})^2\;t^2}\cdot \int d\Omega_{ll}^{CM_{l^+l^-}} L_{\mu\nu}H^{\mu\nu}.
\end{equation}
The kinematical invariant $t$ is in one-to-one relation with the recoiling proton lab momentum $\vec{p^\prime}$ (or energy $E^{\prime}$):
\begin{align}\label{TauLab}
|\vec{p^\prime}|&=2M\sqrt{\tau(1+\tau)},\\
E^{\prime}&=M(1+2\tau),
\end{align}
whereas the invariant $s_{ll}$ is then determined from the recoiling proton lab scattering angle:
\begin{equation}\label{ScatterAngleLab}
\text{cos}\;\theta_{p^{\prime}}=\frac{s_{ll}+2(s+M^2)\tau}{2(s-M^2)\sqrt{\tau(1+\tau)}},
\end{equation}
where $s$ is the squared center-of-mass energy, which can be expressed in terms of the initial photon-beam energy $E_{\gamma}$:
\begin{equation}
s=2E_{\gamma}M+M^2.
\end{equation}

In Ref. \cite{Pauk:2015oaa}, the authors calculated the ratio $R$ of cross sections between electron- and muon-pair production:
\begin{equation}
R(s_{ll},s^0_{ll})\equiv\frac{\left[\sigma_0(\mu^+\mu^-)\right](s_{ll})+[\sigma_0(e^+e^-)](s_{ll})}{[\sigma_0(e^+e^-)](s^0_{ll})},\label{ratio-def}
\end{equation}
which depends on the invariant mass of the lepton pair $s_{ll}$, and a reference point $s_{ll}^0$ to which the measurement is normalized. 

\begin{figure}
	\includegraphics[scale=0.9]{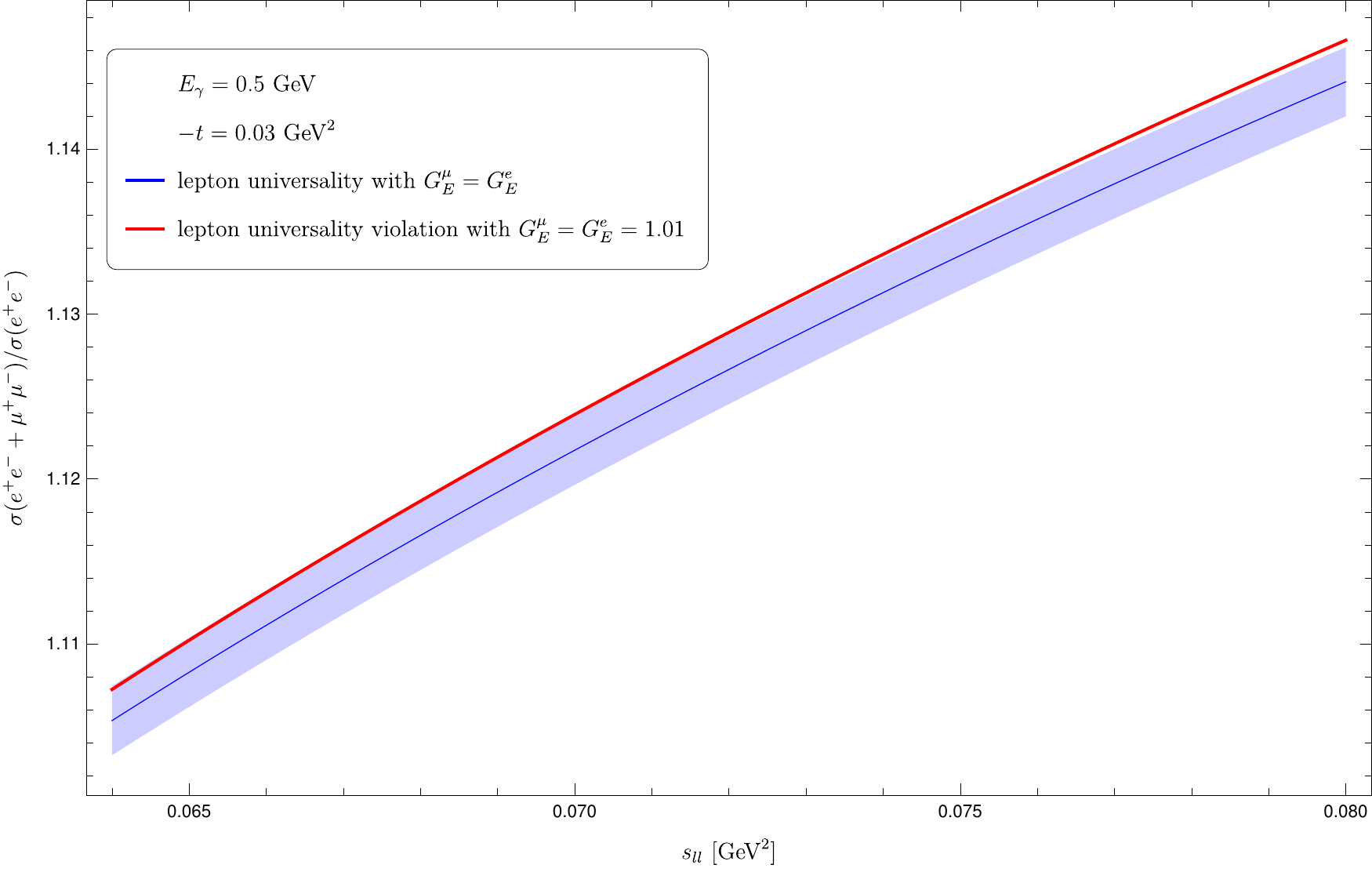}\hspace{0.cm}
	\caption{Ratio of the cross sections in $\gamma p\rightarrow (e^+e^-+\mu^+\mu^-)p$ vs $\gamma p\rightarrow (e^+e^-)p$. The blue band corresponds to a $3\sigma$ band, where $\sigma=7\times10^{-4}$.\label{RatioTree}}
\end{figure}

The corresponding plot for the kinematical range accessible at MAMI is shown in Fig. \ref{RatioTree}. The normalization is shown for the choice $s_{ll}^0=s_{ll}$, i.e., at each point above the muon-pair production threshold the sum of the cross sections for muon- and electron-pair production is divided by the corresponding cross section for electron-pair production. In this plot, the blue curve describes the scenario, when lepton universality holds, i.e., $G_E^{\mu}=G_E^{e}$, while the red curve corresponds to a case when lepton universality is broken by an amount of $1\%$. The blue band describes the $3\sigma$ deviation if this observable is measured with an absolute accuracy of $7\times 10^{-4}$. We will show in this work that radiative corrections shift this curve by more than $3\sigma$, making their inclusion indispensable for a comparison with experiment.

\label{sec2}
\section{Leading-order radiative corrections in the soft-photon limit}
\label{sec3}
We evaluate the first-order QED corrections to the $\gamma p\rightarrow l^+l^-p$ process in the soft-photon limit. This limit is defined by a scaling of the momenta $k$ of virtual photons in the loops and real photon momenta in the bremsstrahlung process, with respect to external scales, as
\begin{equation}
k\sim\lambda,
\end{equation}
where $\lambda$ is a small parameter. We calculate the diagrams at leading order in $\lambda$. This procedure reproduces all infrared-divergent contributions and results in a finite, gauge-invariant piece. The resulting cross-section correction factorizes in terms of the tree-level cross section given by Eq. \eqref{CrossSection}.
\subsection{Virtual corrections}
We start by calculating the one-loop virtual radiative corrections. In the soft-photon approximation, only box diagrams contribute. We list all propagators and their scaling with $\lambda$ in Tab. \ref{PropScale}:
\begin{center}
\begin{table}[H]
\centering
\begin{tabular}{|c|c|}
\hline 
propagator denominator &  scaling at least as\\
\hline 
$(k+p_3)^2-m^2$ &  $\lambda$ \\ 
$(k-p_4)^2-m^2$ &   $\lambda$ \\ 
$k^2$&  $\lambda^2$\\
$(p_3-p_1+k)^2-m^2$ & $1$\\
\hline 
\end{tabular} 
\caption{Scaling of the propagator denominators with the expansion parameter $\lambda$. Only integrals with the first $3$ propagators contribute, since these integrals have a denominator scaling as $\lambda^4$, which is the scaling of
 the integral measure in the numerator. \label{PropScale}}
\end{table}
\end{center}

The integral measure $d^4k$ scales as
\begin{equation}
d^4k\sim \lambda^4.
\end{equation}
Therefore, to obtain a contribution of order $1$, we need a denominator of order $\lambda^4$. This is only possible for the box diagrams when the first $3$ propagators of Tab. \ref{PropScale} are present in a Feynman integral.

\begin{figure}
	\hspace{1.cm} \includegraphics[scale=0.70]{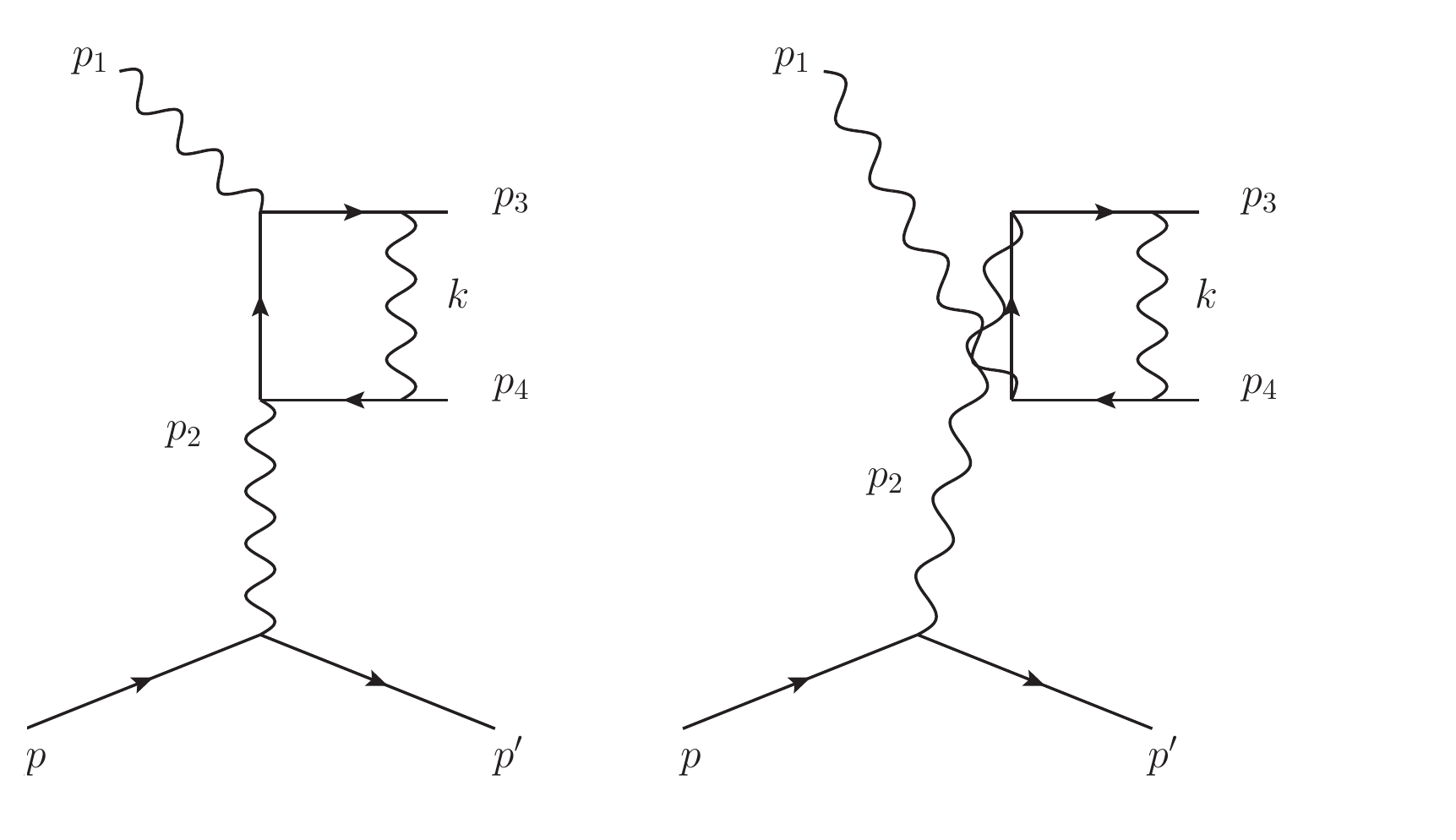}
	\caption{QED box diagrams contributing to the radiative corrections calculation in the soft-photon approximation.\label{softBox}}
\end{figure}
For the box diagrams shown in Fig. \ref{softBox}, we obtain the following leading contribution:
\begin{align}
&\mathcal{M}^\text{box}=(ie^2)\;4\cdot (p_3 p_4)\cdot \mathcal{M}_0\;\mu^{4-d}\int\frac{d^dk}{(2\pi)^d}\frac{1}{(p_3+k)^2-m^2}\frac{1}{(k-p_4)^2-m^2}\frac{1}{k^2}+\mathcal{O}(\lambda)\nonumber\\
&=-\frac{e^2}{8\pi^2}\left(s_{ll}-2m^2\right)\cdot \mathcal{M}_0\cdot C_0\left(m^2, s_{ll}, m^2,0, m^2, m^2\right),\label{BoxVirtual}
\end{align}
with the 3-point function $C_0$ in dimensional regularization, see Ref. \cite{Beenakker:2002nc}:\footnote{We use the same notation for this function as in \url{http://qcdloop.fnal.gov/}}
\begin{align}
&C_0\left(m^2, s_{ll}, m^2,0, m^2, m^2\right)=\frac{1}{s_{ll}\beta}\left\{\left[\frac{1}{\epsilon_\text{IR}}-\gamma_E+\ln\left(\frac{4\pi\mu^2}{m^2}\right)\right]\ln \left(\frac{\beta-1}{\beta+1}\right)\right.\nonumber\\
&\left.+2\;\text{Li}_2\left(\frac{\beta-1}{2\beta}\right)+\ln^2\left(\frac{\beta-1}{2\beta}\right)-\frac{1}{2}\ln^2\left(\frac{\beta-1}{\beta+1}\right)-\frac{\pi^2}{6}\right\}.\label{CFunct}
\end{align}

In Eqs. \eqref{BoxVirtual}, \eqref{CFunct}, $\mu$ is a scale introduced to account for the correct energy dimension of the integral. Physical quantities have to be independent of this scale, as well as of the infrared regulator $\epsilon_\text{IR} \equiv 2-d/2<0$.
All other diagrams are infrared finite and scale at least as $\lambda$. Therefore, the other graphs do not contribute in the soft-photon limit.

\begin{figure}
	\includegraphics[scale=0.7]{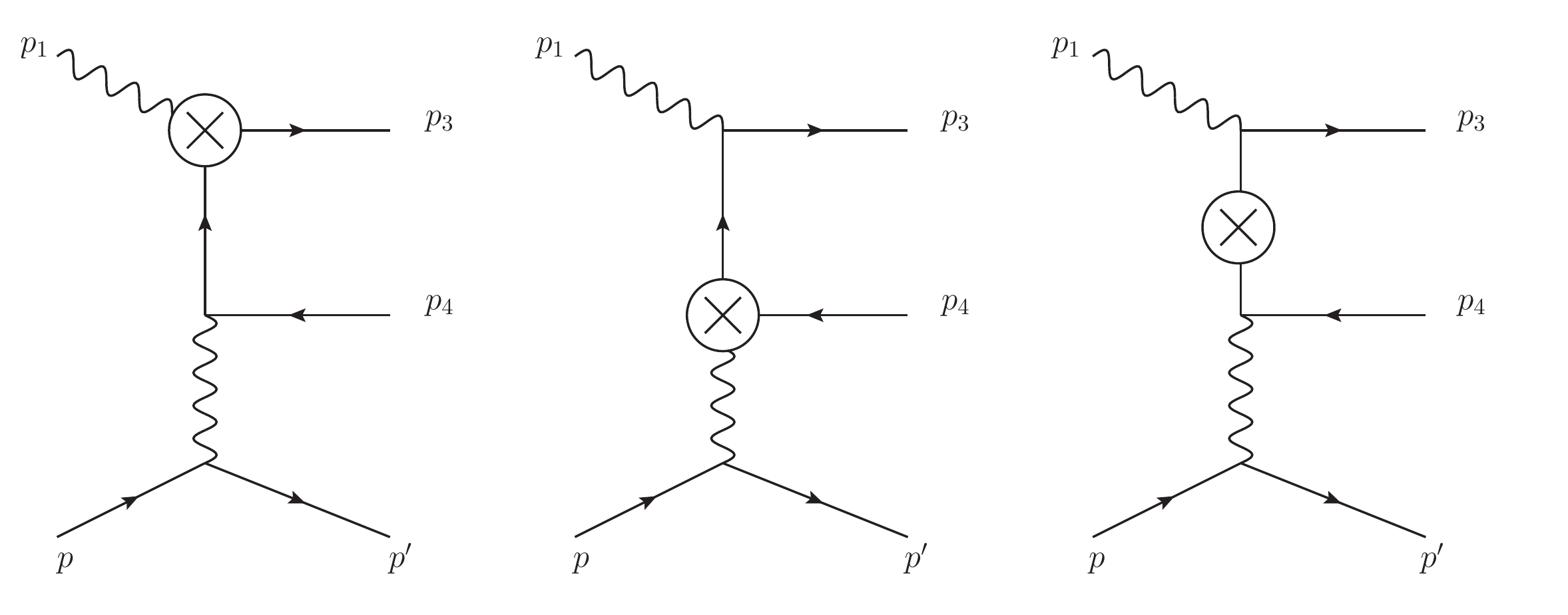}\hspace{0.cm}
	\caption{Counterterm diagrams, which contribute to the $\gamma p\rightarrow l^+l^-p$ process. These give rise to infrared-divergent contributions in the on-shell subtraction scheme and have therefore to be accounted for when calculating the radiative corrections in the soft-photon approximation.\label{CT-Top}}
\end{figure}

Although the box diagrams are UV finite, we have to include counterterm corrections, shown in Fig. \ref{CT-Top}, since they contain infrared-divergent parts in the on-shell subtraction scheme, which we follow here. We describe these contributions according to Ref. \cite{Vanderhaeghen:2000ws}.

\begin{figure}
\includegraphics[trim =1cm 9cm 0cm 9cm, clip=true ,scale=0.7]{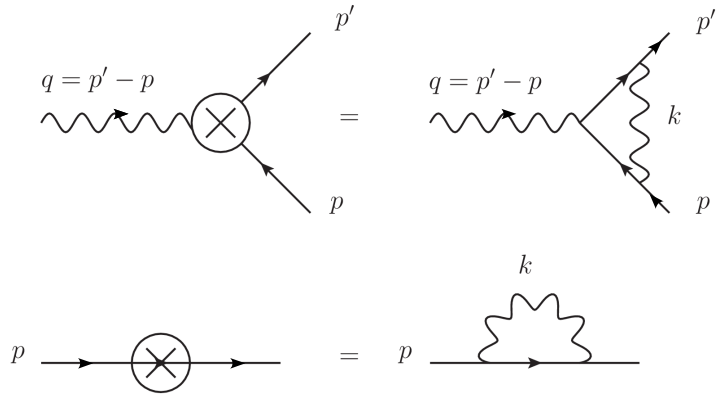}
	\caption{Diagrams for the calculation of the counterterms. The upper diagram defines the vertex counterterm, the lower diagram corresponds to the lepton self-energy\label{VertexCT}}
\end{figure}

In the on-shell subtraction scheme, the vertex counterterm is defined to fix the electron charge $e$ at $q^2=0$. Considering the vertex function in Fig. \ref{VertexCT}, one can decompose the diagram into two tensor structures with corresponding form factors $F$ and $G$:
\begin{align}
\bar{u}(p^{\prime})\Gamma^{\mu}u(p)=\bar{u}(p^{\prime})\left[(1+F(q^2))\gamma^{\mu}+iG(q^2)\sigma^{\mu\nu}\frac{q_{\nu}}{2m}\right]u(p),
\end{align} 
with 
\begin{equation}
q=p^{\prime}-p.
\end{equation}
Only $F(q^2)$ is UV divergent, and one finds at $q^2=0$ the renormalization constant:
\begin{align}
Z_1&=1-F(0)=\nonumber\\
&=1-\frac{e^2}{(4\pi)^2}\left\{\left[\frac{1}{\epsilon_\text{UV}}-\gamma_E+\ln\left(\frac{4\pi\mu^2}{m^2}\right)\right]+2\left[\frac{1}{\epsilon_\text{IR}}-\gamma_E+\ln\left(\frac{4\pi\mu^2}{m^2}\right)\right]+4\right\}.
\end{align}
This leads to the renormalized vertex:
\begin{equation}
\tilde{\Gamma}^{\mu}=\Gamma^{\mu}+(Z_1-1)\gamma^{\mu},
\end{equation}
that in the soft-photon limit $(\tilde{\Gamma}^\mu_s)$, which corresponds to taking only the infrared-divergent part, is expressed as
\begin{equation}
\tilde{\Gamma}_{\text{s}}^{\mu}=-\frac{\alpha}{2\pi}\gamma^{\mu}\left[\frac{1}{\epsilon_\text{IR}}-\gamma_E+\ln\left(\frac{4\pi\mu^2}{m^2}\right)\right].
\end{equation}
The contribution of the two vertex counterterms in Fig. \ref{CT-Top} is then given by
\begin{equation}
\mathcal{M}_\text{vertex}^{\text{ct}}=-\frac{\alpha}{\pi}\left[\frac{1}{\epsilon_\text{IR}}-\gamma_E+\ln\left(\frac{4\pi\mu^2}{m^2}\right)\right]\mathcal{M}_0.\label{CTVertex}
\end{equation}

The self-energy counterterm is defined from the lepton self-energy correction $\Sigma(p)$, which is expressed in terms of the lepton propagator $S$:
\begin{equation}
iS=iS^0+iS^0(-i)\Sigma(p)iS,
\end{equation}
with free fermion propagator given by
\begin{equation}
S^0(p)=\frac{\sla{p}+m}{p^2-m^2}.
\end{equation}  
Calculating up to first order, we have to include the one-loop correction:
\begin{equation}
-i\Sigma(\sla{p})=-e^2\mu^{4-d}\int\frac{d^dk}{(2\pi)^d}\frac{\gamma^{\alpha}(\sla{p}+\sla{k}+m)\gamma_{\alpha}}{((p+k)^2-m^2)\;k^2}.
\end{equation}
The on-shell renormalization condition fixes the pole at $p^2=m^2$ with residue equal to one. This gives the renormalization constants $Z_2$ and $Z_m$:
\begin{align}
Z_2&=1+\left.\frac{d\;\Sigma (\sla{p})}{d\sla{p}}\right|_{\sla{\;p}\;=m},\\
(1-Z_m)Z_2m&=\Sigma(m).
\end{align}
The evaluation of $\Sigma(p)$ and its derivative, results in the renormalization constants:
\begin{align}
Z_2&=1-\frac{e^2}{(4\pi)^2}\left\{\left[\frac{1}{\epsilon_\text{UV}}-\gamma_E+\ln\left(\frac{4\pi\mu^2}{m^2}\right)\right]+2\left[\frac{1}{\epsilon_\text{IR}}-\gamma_E+\ln\left(\frac{4\pi\mu^2}{m^2}\right)\right]+4\right\},\\
Z_2Z_m&=1-\frac{e^2}{(4\pi)^2}\left\{4\left[\frac{1}{\epsilon_\text{UV}}-\gamma_E+\ln\left(\frac{4\pi\mu^2}{m^2}\right)\right]+2\left[\frac{1}{\epsilon_\text{IR}}-\gamma_E+\ln\left(\frac{4\pi\mu^2}{m^2}\right)\right]+8\right\}.
\end{align}
The renormalized self-energy is then given by
\begin{equation}
\tilde{\Sigma}(p)=\Sigma(p)-(Z_2-1)\sla{p}+(Z_2Z_m-1)m.
\end{equation}
Taking only the infrared-divergent piece in the soft-photon limit $(\tilde{\Sigma}_\text{s})$, we obtain:
\begin{equation}
\tilde{\Sigma}_\text{s}(p)=\frac{\alpha}{2\pi}(\sla{p}-m)\left[\frac{1}{\epsilon_\text{IR}}-\gamma_E+\ln\left(\frac{4\pi\mu^2}{m^2}\right)\right].
\end{equation}
The contribution of the self-energy counterterm $M^\text{ct}_\text{se}$ in Fig. \ref{CT-Top} is therefore given by
\begin{equation}
\mathcal{M}_{\text{se}}^{\text{ct}}=\frac{\alpha}{2\pi}\left[\frac{1}{\epsilon_\text{IR}}-\gamma_E+\ln\left(\frac{4\pi\mu^2}{m^2}\right)\right]\mathcal{M}_0.\label{CTSE}
\end{equation}

Adding virtual corrections of Eq. \eqref{BoxVirtual} and counterterms of Eqs. \eqref{CTVertex} and \eqref{CTSE}, we obtain the virtual one-loop correction in the soft-photon limit $\mathcal{M}_{\text{s;V}}$:
\begin{align}
\mathcal{M}_{\text{s;V}}=-\frac{\alpha}{2\pi}\left\{\left(s_{ll}-2m^2\right) C_0\left(m^2, s_{ll}, m^2,0, m^2, m^2\right)+\left[\frac{1}{\epsilon_\text{IR}}-\gamma_E+\ln\left(\frac{4\pi\mu^2}{m^2}\right)\right] \right\}\mathcal{M}^0.
\end{align}

The resulting virtual correction to the cross section is then given, to first order in $\alpha$, by
\begin{equation}
\left(\frac{d\sigma}{dt_{ll}ds_{ll}}\right)_{\text{s;V}}=2\;\text{Re}\left[\mathcal{M}_0^*\times\mathcal{M}_{\text{s;V}}\right].
\end{equation}
It can be expressed as
\begin{equation}
\left(\frac{d\sigma}{dt_{ll}ds_{ll}}\right)_{\text{s;V}}=\left(\frac{d\sigma}{dt_{ll}ds_{ll}}\right)_{0}\left(\vphantom{\frac{1}{2}}\delta^\text{IR}_\text{s;V}+\delta_\text{s;V}\right),
\end{equation}
with the infrared-divergent part:
\begin{equation}
\delta^\text{IR}_\text{s;V}=\left(\frac{-\alpha}{\pi}\right)\left[\left(\frac{1+\beta^2}{2\beta}\right)\ln\left(\frac{1-\beta}{1+\beta}\right)+1\right]\left[\frac{1}{\epsilon_\text{IR}}-\gamma_E+\ln\left(\frac{4\pi\mu^2}{m^2}\right)\right],\label{IRVirtual}
\end{equation}
and the finite part:
\begin{align}
\delta_\text{s;V}=&\left(\frac{-\alpha}{\pi}\right)\left(\frac{1+\beta^2}{2\beta}\right)
\left\{2\;\text{Li}_2\left(\frac{2\beta}{\beta+1}\right)+\frac{1}{2}\ln^2\left(\frac{1-\beta}{1+\beta}\right)-\pi^2\right\}.\label{FiniteVirtual}
\end{align}
\subsection{Soft-photon bremsstrahlung}
\begin{figure}
	\includegraphics[scale=0.70]{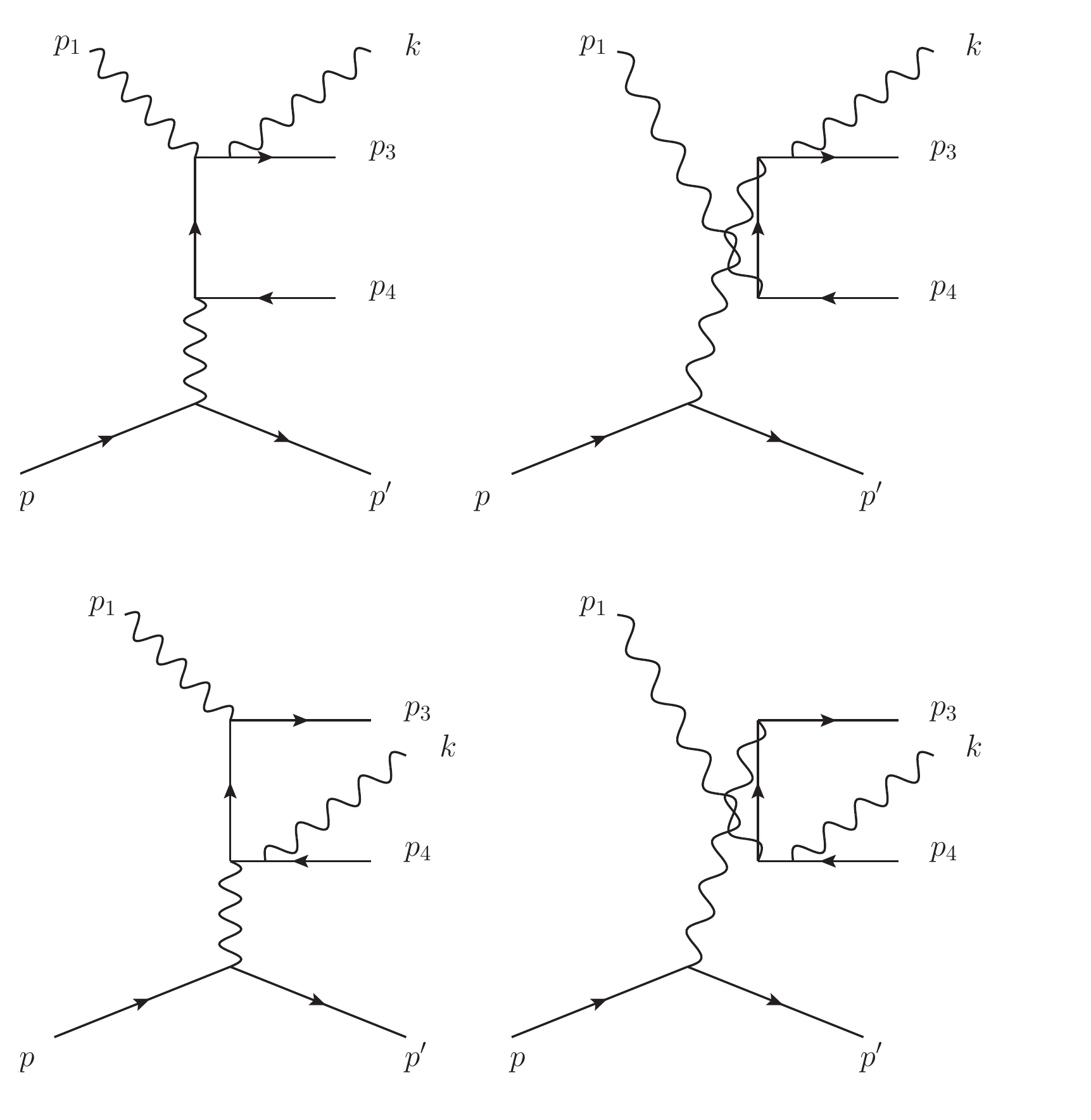}\hspace{0.cm}
	\caption{Diagrams with real photon emission from the lepton lines for the Bethe-Heitler process. In the soft-photon limit, the diagram with the photon attached to the internal (off-shell) fermion line does not contribute.\label{SoftBrems}}
\end{figure}
Besides the QED virtual radiative corrections, one has to account for processes with radiation of undetected photons.

The diagrams contributing to the soft bremsstrahlung from the lepton side are shown in Fig. \ref{SoftBrems}. Note that the diagram, where the photon is attached to the internal lepton line, vanishes for $\lambda\rightarrow 0$ and does not contribute in the soft-photon limit. Denoting the momentum of the photon by $k$, we find the squared matrix element for this process in the form:
\begin{align}
&\left|\mathcal{M}(\gamma p\rightarrow\gamma_\text{s}\;l^+l^-p)\right|^2=\left|\mathcal{M}(\gamma p\rightarrow l^+l^-p)\right|^2\;(-e^2)\left[ \frac{p_3^{\mu}}{p_3\cdot k}-\frac{p_4^{\mu}}{p_4\cdot k}\right]\cdot\left[ \frac{p_{3\mu}}{p_3\cdot k}-\frac{p_{4\mu}}{p_4\cdot k}\right].
\end{align}
To calculate the contribution to the cross section, one then has to integrate over the undetected soft-photon energy up to a small value $\Delta E_s$, determined by the experimental resolution. 

Due to the energy-momentum conserving $\delta$-function, $\delta^4(p_1+p-p_3-p_4-p^{\prime}-k)$, the integration domain has a complicated shape in the lab system. The integration can be carried out in the rest frame $\mathcal{S}$ of the real ($p_1)$ and virtual ($p_2$) photons, which is also the rest frame of the di-lepton pair and soft photon, defined by
\begin{equation}
\vec{p}_1+\vec{p}_2=\vec{p}_3+\vec{p}_4+\vec{k}=0.
\end{equation}
In such frame, the dependence of the integral with respect to the soft-photon momentum becomes isotropic.
For the differential cross section, we then need to evaluate:
\begin{equation}
\left(\frac{d\sigma}{dt ds_{ll}}\right)_\text{s;R}=-\left(\frac{d\sigma}{dt ds_{ll}}\right)_0\frac{e^2}{(2\pi)^3}\int_{|\vec{k}|<\Delta E_s}\frac{d^3\vec{k}}{2k^0}\left[\frac{m^2}{(p_3k)^2}+\frac{m^2}{(p_4k)^2}-\frac{2(p_3p_4)}{(p_3k)(p_4k)}\right ],\label{SoftPhotonIntegral}
\end{equation}
where the integration is performed in the frame $\mathcal{S}$. 

The integrals are infrared divergent and can be carried out analytically after dimensional regularization. They have been worked out, e.g., in Ref. \cite{tHooft:1978jhc}.
For the kinematics in system $\mathcal{S}$, where the soft-photon momentum:
\begin{equation}
|\vec{k}|\ll\left|\vec{p}_3\right|,\left|\vec{p}_4\right|,
\end{equation}
with the lepton momenta:
\begin{align}
p_3^0=p_4^0=\frac{\sqrt{s_{ll}}}{2},\ \ \vec{p}_3=-\vec{p}_4,
\end{align}
we obtain:
\begin{align}
\left(\frac{d\sigma}{dt ds_{ll}}\right)_\text{s;R}&=\left(\frac{d\sigma}{dtds_{ll}}\right)_{0}\left(\vphantom{\frac{1}{2}}\delta_\text{s;R}^\text{IR}+\delta_\text{s;R}\right),
\end{align}
where $\delta_\text{s;R}^\text{IR}$ is the infrared-divergent contribution due to real photon emission:
\begin{align}
\delta^\text{IR}_\text{s;R}=\left(\frac{-\alpha}{\pi}\right)\left[\left(\frac{1+\beta^2}{2\beta}\right)\ln\left(\frac{1+\beta}{1-\beta}\right)-1\right]\left[\frac{1}{\epsilon_\text{IR}}-\gamma_E+\ln\left(\frac{4\pi\mu^2}{m^2}\right)\right], \label{IRReal}
\end{align}
and $\delta_\text{s;R}$ is the corresponding finite part:
\begin{align}
\delta_\text{s;R}=\left(\frac{-\alpha}{\pi}\right)\left\{\ln\left(\frac{4\Delta E_s^2}{m^2}\right)\right.&\left.\left[1+\left(\frac{1+\beta^2}{2\beta}\right)\ln\left(\frac{1-\beta}{1+\beta}\right)\right]+\frac{1}{\beta}\ln\left(\frac{1-\beta}{1+\beta}\right)+\right.\nonumber\\
\phantom{=}+\left(\frac{1+\beta^2}{2\beta}\right)&\left.\left[2\;\text{Li}_2\left(\frac{2\beta}{1+\beta}\right)+\frac{1}{2}\ln^2\left(\frac{1-\beta}{1+\beta}\right)\right]\right\}.\label{FiniteReal}
\end{align}

The maximum value of the undetected soft-photon energy $\Delta E_s$ is defined in the system $\mathcal{S}$. One can re-express it in terms of the detector resolutions. We consider the case of detecting the recoil proton only. The energy $E^{\prime}$ and angle $\theta_{p^{\prime}}$ of the scattered proton are measured in the lab frame.
The missing mass $ M_\text{miss} $ of the system is defined by
\ber
M^2_\text{miss} & = & (p_3 + p_4 + k )^2 = s_{ll} + 2 M_\text{miss}  E_s,  \\
E_s & = & \frac{M^2_\text{miss} - s_{ll}}{2 M_\text{miss}},\label{Reso}
\eer
where $E_s$ denotes the soft-photon energy.

The missing mass $M_\text{miss}$ is experimentally determined from the quantity:
\begin{align}
M_{\text{miss}}^2&=(p_1+p-p^{\prime})^2 \nonumber\\
&=4M\tau \left(E_{\gamma}\sqrt{\frac{1+\tau}{\tau}}\text{cos}\;\theta_{p^{\prime}}-E_{\gamma}-M\right),\label{MissMassLab}
\end{align}
where $\tau$ is determined from the lab proton momentum by Eq. \eqref{TauLab}, and $\theta_{p^{\prime}}$ is the experimentally measured recoil proton scattering angle in the laboratory frame.

For the process without radiation, this angle is given by Eq. \eqref{ScatterAngleLab}, which can be equivalently obtained from Eq. \eqref{MissMassLab} by the replacement $M_{\text{miss}}^2 \rightarrow s_{ll}$:
\begin{equation}
s_{ll}=4M\tau\left( E_{\gamma}\sqrt{\frac{1+\tau}{\tau}}\text{cos}\;\theta_{p^{\prime}}|_{\text{no rad}}-E_{\gamma}-M\right)\label{sllLab}.
\end{equation}

Combining Eqs. \eqref{MissMassLab} and \eqref{sllLab}, we can express the soft-photon energy of Eq. \eqref{Reso} approximately as:
\begin{equation}
E_\text{s}=\frac{2ME_{\gamma}\sqrt{\tau(1+\tau)}}{\sqrt{s_{ll}}}\left[\vphantom{\frac{1}{2}}\text{cos}\;\theta_{p^{\prime}}-\text{cos}\;\theta_{p^{\prime}}|_{\text{no rad}}\right].
\end{equation}

Consequently, the experimental recoiling proton angular resolution, denoted as $\Delta\theta_{p^\prime}$, determines the maximum value $\Delta E_s$ of the undetected soft-photon energy, which enters the radiative correction of Eq. \eqref{FiniteReal}, as
\begin{equation}
\Delta E_s=\frac{2ME_{\gamma}\sqrt{\tau(1+\tau)}}{\sqrt{s_{ll}}}\;\text{sin}\;\theta_{p^{\prime}}\;\Delta\theta_{p^{\prime}}.\label{AngularReso}
\end{equation}

\subsection{Total result and exponentiation}
Adding the real and virtual contributions of Eqs. \eqref{IRReal} and \eqref{IRVirtual}, we find a cancellation of all infrared divergences on the level of the cross section:
\begin{equation}
\delta_{\text{s;R}}^\text{IR}+\delta_{\text{s;V}}^\text{IR}=0.
\end{equation}
For the finite part of the first-order QED corrections in the soft-photon approximation:
\begin{equation}
\delta=\delta_{\text{s;R}}+\delta_{\text{s;V}},
\end{equation} 
we find the result:
\begin{align}
\delta&=-\left(\frac{\alpha}{\pi}\right)\left\{\left[\ln\left(\frac{4\Delta E_s^2}{m^2}\right)+\ln\left(\frac{1-\beta}{1+\beta}\right)\right]\left[1+\left(\frac{1+\beta^2}{2\beta}\right)\ln\left(\frac{1-\beta}{1+\beta}\right)\right]\right.\nonumber\\
&\phantom{=}\left.+\left(\frac{1-\beta}{\beta}\right)\ln\left(\frac{1-\beta}{1+\beta}\right)
+\left(\frac{1+\beta^2}{2\beta}\right)
\left[4\;\text{Li}_2\left(\frac{2\beta}{1+\beta}\right)- \pi^2\right]\right\},
\end{align}
which reduces in the limit $s_{ll}>>4m^2$ to:
\begin{equation}
\delta=-\left(\frac{\alpha}{\pi}\right)\left\{\ln\left(\frac{4\Delta E_s^2}{s_{ll}}\right)\left[1+\ln\left(\frac{m^2}{s_{ll}}\right)\right]-\frac{\pi^2}{3}\right\}.
\end{equation}

To account for the emission of a higher amount of soft photons or higher-order virtual corrections due to soft photons in the loop, we follow Ref. \cite{Yennie:1961ad} and exponentiate the terms leading to double logarithmic enhancements as

\begin{align}
\left(\frac{d\sigma}{dt\;ds_{ll}}\right)_{\text{s;tot}}&=\left(\frac{d\sigma}{dt\;ds_{ll}}\right)_{0}\cdot F\exp\left\{-\frac{\alpha}{\pi}\left[\ln\left(\frac{4\Delta E_s^2}{m^2}\right)+\ln\left(\frac{1-\beta}{1+\beta}\right)\right]\left[1+\left(\frac{1+\beta^2}{2\beta}\right)\ln\left(\frac{1-\beta}{1+\beta}\right)\right]\right\}\nonumber\\
&\times
\left\{1-\frac{\alpha}{\pi}\left[\left(\frac{1-\beta}{\beta}\right)\ln\left(\frac{1-\beta}{1+\beta}\right)
+\left(\frac{1+\beta^2}{2\beta}\right)\left[4\;\text{Li}_2\left(\frac{2\beta}{1+\beta}\right)- \pi^2\right]\right]\right\}\nonumber\\
&\equiv\left(\frac{d\sigma}{dt\;ds_{ll}}\right)_{0}(1+\delta_\text{exp}).
\label{Exponentiated}
\end{align}

Note that in Eq. \eqref{Exponentiated} terms of single logarithmic nature of order $\alpha$ are still missing, and require a full one-loop calculation. The normalization factor $F$ in Eq. \eqref{Exponentiated} is due to the physical assumption that in an  experiment the sum of all soft-photon energies is smaller than $\Delta E_s$, instead of requiring that each soft-photon energy is individually smaller than $\Delta E_s$. It was shown in Ref. \cite{Yennie:1961ad} that when including the leading correction from unity, the normalization factor $F$ is given by:
\begin{equation}
F=1-\frac{\alpha^2}{3}\left[1+\left(\frac{1+\beta^2}{2\beta}\right)\ln\left(\frac{1-\beta}{1+\beta}\right)\right]^2+...
\end{equation}
Although we account for the factor $F$ explicitly, its deviation from unity is quite small: for $s_{ll}=0.077\text{ GeV}^2$ approximately $-2.4 \times 10^{-3}$ for electrons and $-8.5 \times 10^{-6}$ for muons.
\newpage
\section{Results and discussion}
\label{sec4}

\begin{figure}[H]
	\includegraphics[scale=0.95]{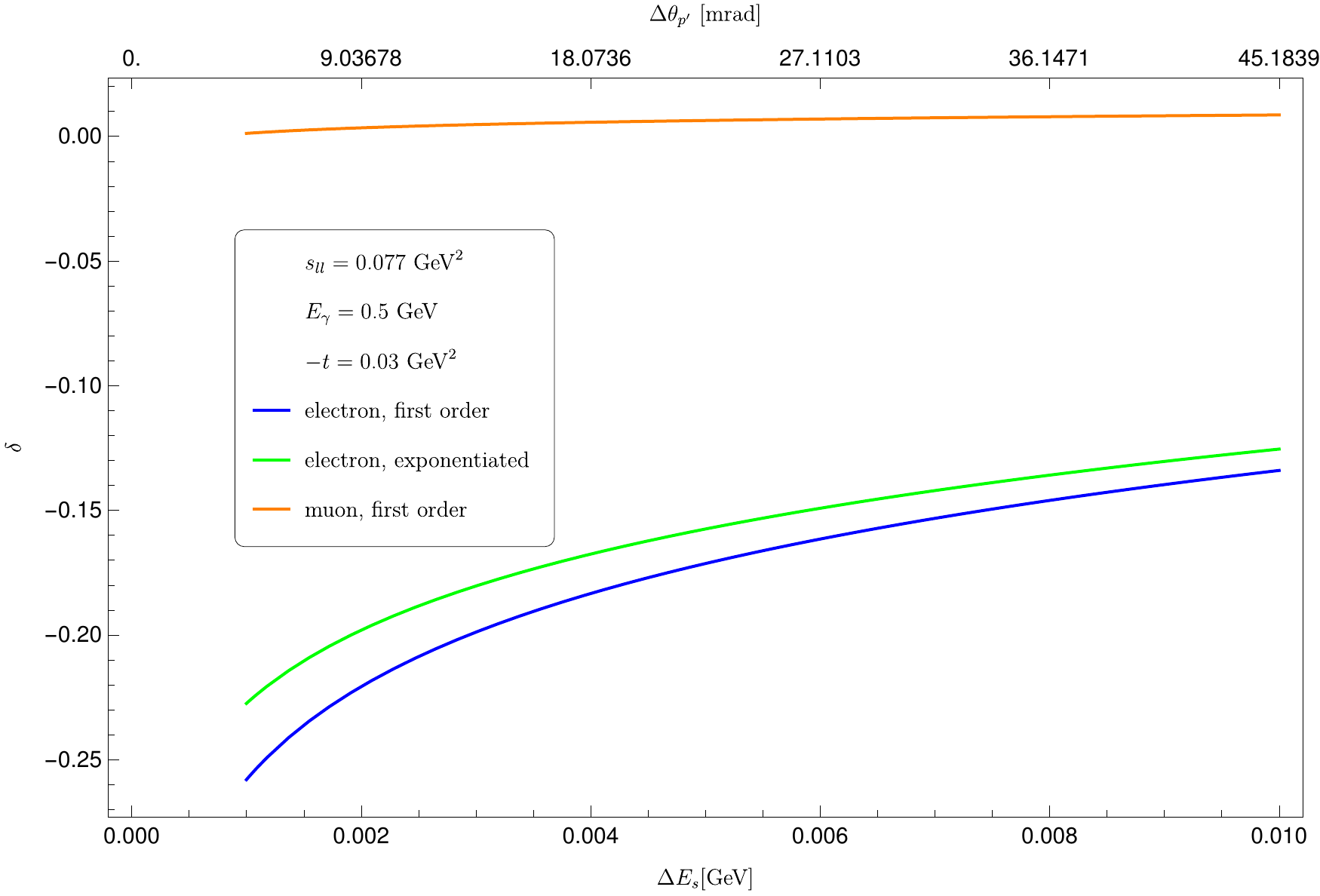}\hspace{0.cm}
	\caption{QED corrections to the cross section in the soft-photon limit as a function of the soft-photon energy $\Delta E_s$, which corresponds to the integrated over angular bins $\Delta\theta_{p^\prime}$ according to Eq. \eqref{AngularReso}. This variation stems from the integrated over radiative tail. The external kinematics and the di-lepton invariant mass $s_{ll}=0.077\ \text{GeV}^2$ are indicated on the plot.}\label{RadTail}
\end{figure}

In Fig. \ref{RadTail}, we show the corrections at fixed $s_{ll}=0.077\text{ GeV}^2$ as a function of the soft-photon energy.
We observe a logarithmic behavior of the correction factor $\delta$ which gives rise to the so-called radiative tail. We also show the exponentiated form, $\delta_\text{exp}$, given by Eq. \eqref{Exponentiated}, which estimates higher-order effects of soft-photon corrections. Assuming a value $\Delta E_\text{s}=0.01\text{ GeV}$, $\delta$ at first order differs by about $0.006$ for electron-pair production and is indistinguishable at the level of precision for muon-pair production (the difference is around $-1.0\times10^{-4}$).

\begin{figure}[H]
	\includegraphics[scale=0.95]{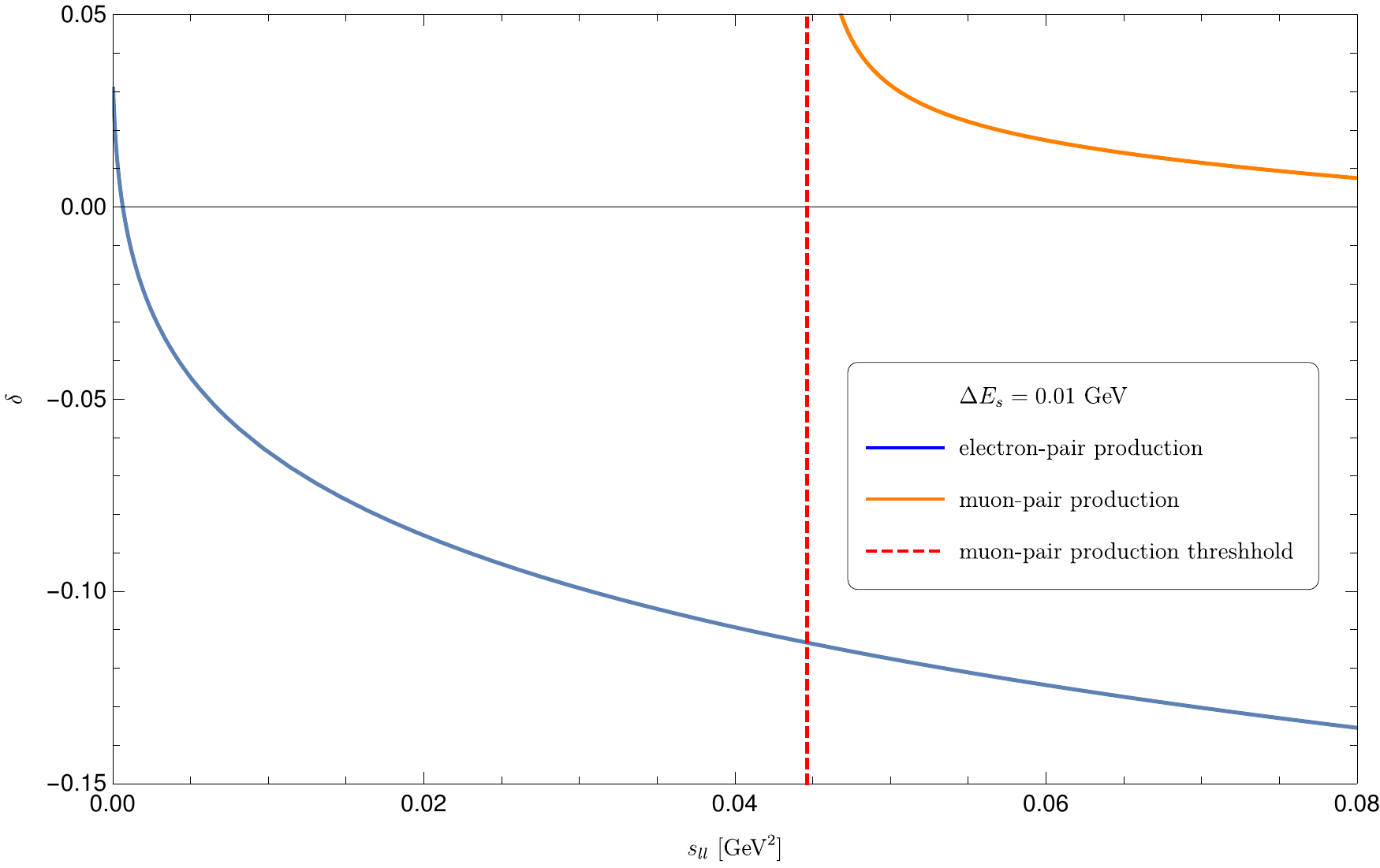}\hspace{0.cm}
	\caption{First-order QED corrections to the cross section in the soft-photon limit, using $\Delta E_s=0.01\text{ GeV}$. The vertical dashed red line indicates the muon-pair production threshold at $s_{ll}\approx 0.045\text{ GeV}^2$.\label{PlotCorrections}}
\end{figure}

In Fig. \ref{PlotCorrections}, we show the radiative corrections to the cross section in the kinematical range of $s_{ll}$ between $0$ and $0.08\ \text{GeV}^2$. The muon threshold is at $s_{ll}=4m_{\mu}^2\approx 0.045\ \text{GeV}^2$ (vertical dashed red line in Fig \ref{PlotCorrections}). We observe that the corrections for electrons are negative of order $10$ percent, while the corrections for muons are positive of order $1$ percent. 

Taking radiative corrections into account, the ratio of Eq. \eqref{ratio-def} is now given by
\begin{equation}
R(s_{ll},s^0_{ll})\equiv\frac{\left[\sigma_0(\mu^+\mu^-)(1+\delta^{\mu})\right](s_{ll})+[\sigma_0(e^+e^-)(1+\delta^e)](s_{ll})}{[\sigma_0(e^+e^-)(1+\delta^e)](s^0_{ll})},
\end{equation}
which depends on the measured invariant lepton mass $s_{ll}$ and the reference point $s^0_{ll}$, to which the cross section is normalized. $\delta^e$ and $\delta^\mu$ are given by Eq. \eqref{Exponentiated}. One chooses $s_{ll}^0<4m_{\mu}^2$, such that the reference measurement is below the muon-pair-production threshold, and only electron pairs are created.

\begin{figure}[H]
	\includegraphics[scale=0.95]{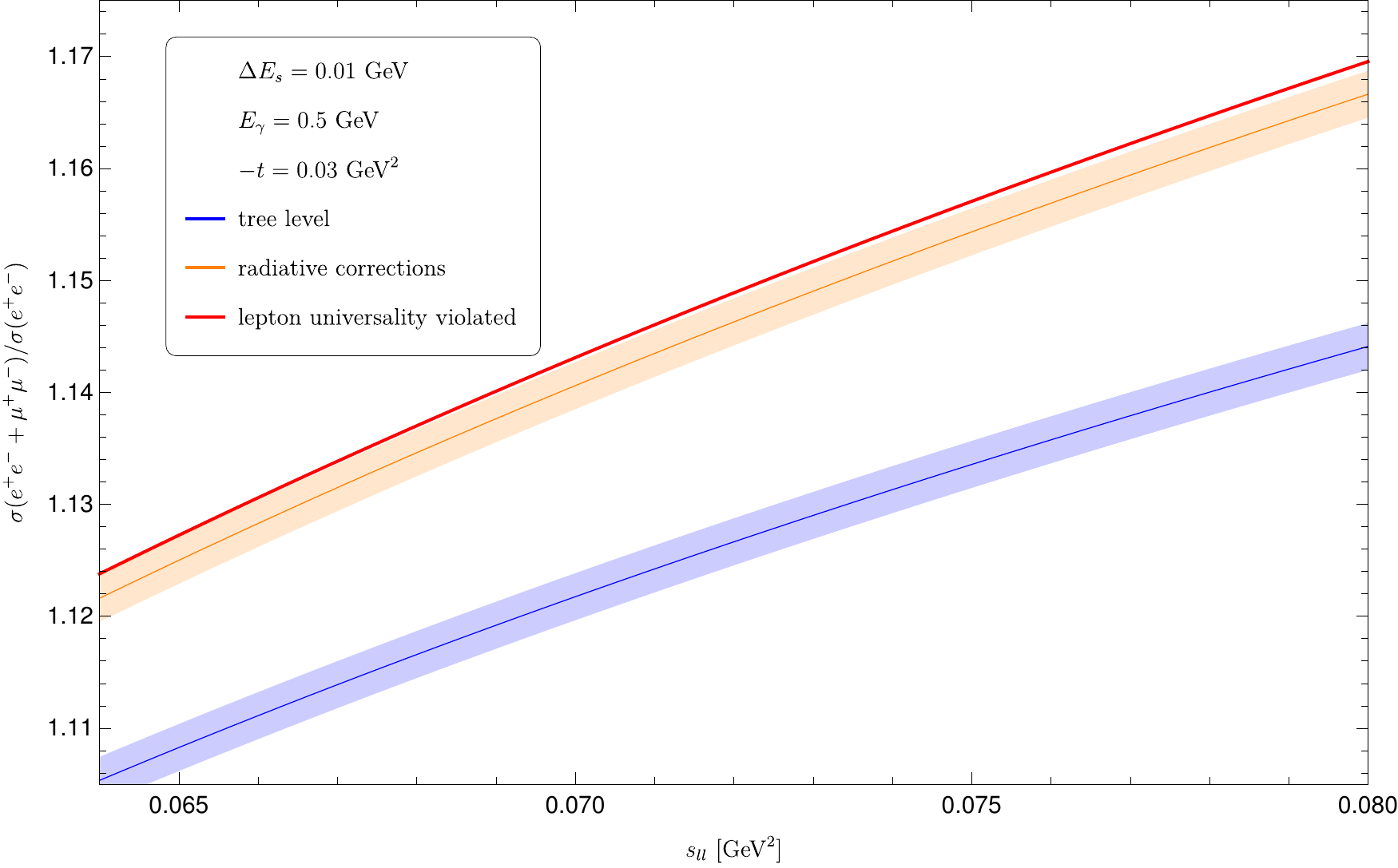}\hspace{0.cm}
	\caption{Ratio of cross sections between electron- and muon-pair production at tree level (blue curve) and with account of first-order QED corrections estimated using $\Delta E_s=0.01 \text{ GeV}$ (orange curve) with $3\sigma$ error bands. The red curve denotes the scenario when lepton universality is broken with $G^{\mu}_E/G^e_E=1.01$, including the radiative corrections in the soft-photon approximation.\label{Ratio1}}
\end{figure}

In Fig. \ref{Ratio1}, we show the differential cross section ratio $R$ of Eq. \eqref{ratio-def}, including first-order QED corrections in the soft-photon approximation with $\Delta E_s=0.01\text{ GeV}$. One sees from this plot, that the inclusion of radiative corrections is indispensable, since the ratio of cross sections, defined in Eq. \eqref{ratio-def}, is shifted to higher values by more than the $3\sigma$ band. The radiative corrections to $R$ are of the order of a few percent. The red curve in Fig. \ref{Ratio1} shows the scenario when lepton universality is violated by $G^{\mu}_E/G^e_E=1.01$. Following Ref. \cite{Pauk:2015oaa}, we use $3\sigma$ bands around the curves, with the experimental resolution $\sigma=7\times 10^{-4}$. The statement that lepton universality can be tested with a $3\sigma$ confidence level remains true if one adds radiative corrections as can be seen in Fig. \ref{Ratio1}.

\begin{figure}
	\includegraphics[scale=0.95]{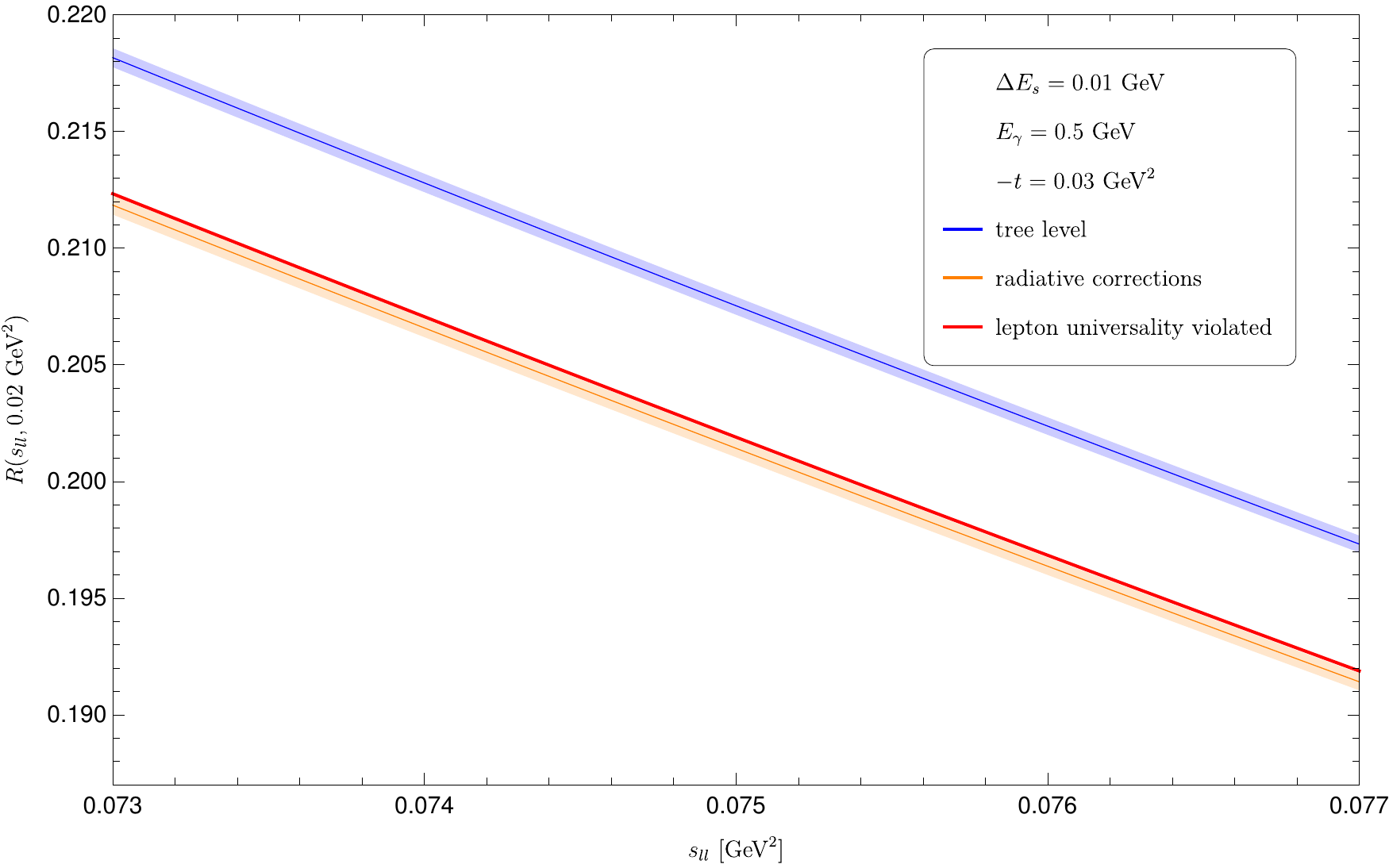}\hspace{0.cm}
	\caption{Ratio of cross sections between electron- and muon-pair production at tree level (blue curve) and with account of first-order QED corrections estimated using $\Delta E_s=0.01 \text{ GeV}$ (orange curve), normalized to the electron-pair production cross section at $s^0_{ll}=0.02\text{ GeV}^2$.\label{Ratio3}}
\end{figure}

In Fig. \ref{Ratio3}, we show the corresponding ratio between the cross sections normalized to a value below the muon-pair production threshold. As a reference point, we choose $s^0_{ll}=0.02\text{ GeV}^2$. The bands now correspond to the renormalized $3\sigma$ bands, i.e., $$\sigma=7\times 10^{-4}\cdot\frac{\sigma(e^+e^-)(s_{ll})}{\sigma(e^+e^-)(s^0_{ll})}.$$

\section{Conclusions and outlook}
\label{sec5}
In this work, we have calculated QED radiative corrections to the photoproduction of electron and muon pairs on a proton target in the soft-photon approximation. Only radiation from the produced pair and box diagrams with photon and lepton legs contribute in this approximation when accounting for the finite lepton mass. The resulting correction to the cross section factorizes in terms of the tree-level contribution. We expressed the proportionality factor in a compact analytical form. With account of radiative corrections, the ratio of photoproduction cross sections of $e^+ e^- + \mu^+ \mu^-$ to $e^+ e^- $ pairs at the same beam energy (as well when compared to the $e^+e^-$ cross section at an energy below the muon-production threshold) increases by a percent amount comparing to the tree level result. Such changes are significantly larger than the precision needed to distinguish between the proton charge radii extractions from experiments with muons and electrons. It makes a correct inclusion of radiative corrections paramount for the experimental realization. As a next step, we plan to extend the radiative correction result in the soft-photon approximation presented in this work to a full one-loop QED calculation on the lepton side and to include the box diagrams resulting from the two-photon exchange between lepton and proton with an intermediate proton state using the techniques developed in Refs. \cite{Tomalak:2014sva,Tomalak:2014dja} for elastic $l^-p$ scattering. For the leading corrections resulting from the hadronic side we expect, from the corresponding results for the elastic $l^- p$ scattering, to receive cross section corrections at the percent level for the electron case. Such anticipated corrections would translate in a change of the ratio of $e^+ e^- + \mu^+ \mu^-$ to $e^+ e^- $ cross sections at the per mille level, corresponding with the 1$\sigma$ accuracy goal discussed above for this quantity. 
\label{sec6}

\section*{Acknowledgements}

We  would like to thank Dr. Aleksandrs Aleksejevs and Shihao Wu for useful discussions.
This work was supported by the Deutsche Forschungsgemeinschaft DFG in part
 through the Collaborative Research Center [The Low-Energy Frontier of the 
Standard Model (SFB 1044)], and in part through the Cluster of Excellence 
[Precision Physics, Fundamental Interactions and Structure of Matter (PRISMA)]. Matthias Heller is supported in part by GRK Symmetry Breaking (DFG/GRK 1581). Our figures were generated using \texttt{Jaxodraw}\cite{Binosi:2003yf}, based on \texttt{AxoDraw} \cite{Vermaseren:1994je}. For our \texttt{Mathematica} plots, we use the package \texttt{MaTeX} \footnote{https://github.com/szhorvat/MaTeX}.

\end{document}